\documentstyle[preprint,aps]{revtex}

\begin{document}
\draft
\preprint{RU9423}
\title{A Fully Isotopic Model of Fragmentation}
\author{K. C. Chase and A. Z. Mekjian}
\address{Department of Physics, Rutgers University \\
Piscataway, New Jersey 08854}
\date{\today}

\maketitle

\begin{abstract}
A general model for the fragmentation of a two-component system
(e.g. protons and neutrons) is proposed and solved exactly.
The extension of this model to any number of components is
also shown to be exactly solvable.
A connection between this models and the permutation
group is discussed.  The notion of isotopic equivalence is defined in
order to evaluate the equivalence of these models to earlier
one-component models.  All the one-component models considered
in earlier papers are shown to be equivalent to a particular
class of two-component models.  A simplified model applicable to the
case of nuclear fragmentation is introduced and analyzed.
Modifications to this model to include effects such as pairing and
Coulomb interactions are discussed.
\end{abstract}


\narrowtext

\section{Introduction}
\label{sec:Introduction}

The fragmentation of mesoscopic systems, particularly heavy nuclear
systems, has attracted considerable interest over the past several
years~\cite{Mekjian1,Mekjian2,Mekjian3,Chase,Sneppen,Jaqaman1,Jaqaman2,DeAngelis1,DeAngelis2,Gross1,Gross2,Gross3,Sobotka,Aichelin,Canupi,Desbois,Koonin,Csernai,Boal,Cole,Fai,Friedman,Bondorf}.
A number of statistical models have been proposed to explain
the phenomena.
In these models, the final state of the fragmentation
of a nucleus of $A$ nucleons is represented by the vector
$\vec{n} = (n_{1}, \ldots, n_{A})$, where
$n_{j}$ is the number of fragments with $j$ nucleons.
The set of all such states or partitions of $A$ is denoted by
$\Pi_{A}$, each of which must satisfy
$\sum_{j=1}^{A} j n_{j} = A$.

In actuality, the final state of a nuclear fragmentation process
is a more complicated object.  Fragments are formed with varying
numbers of neutrons and protons, which are distinguishable objects.
For this reason, a representation of the final fragmentation state
of such a two-component or isotopic system should minimally
contain information on the neutron and proton content of
each fragment.  In this representation, the final state of the
fragmentation of a nucleus containing $Z$ protons and $N$ neutrons
is given by a vector of $(Z+1)(N+1)-1$ integers,
$\vec{n} = (n_{0 1}, \ldots, n_{0 N}, n_{1 0}, \ldots,
n_{1 N}, \ldots, n_{Z N})$,
where $n_{j k}$ is the number of fragments with $j$
protons and $k$ neutrons.
The set of all such states or partitions of $A=Z+N$ is denoted by
$\Pi_{ZN}$, each of which is subject to the two constraints
$\sum_{j k} j n_{j k} = Z$, $\sum_{j k} k n_{j k} = N$.

In general the fragmentation of an $N$-component system of
$A = \sum_{j=1}^{N} A_{j}$ particles where $A_{j}$ is the number
of indistinguishable particles of type $j$ can be described by
a vector of $\prod_{j=1}^{N} (A_{j}+1) - 1$ integers,
$\vec{n} = (n_{00...01}, \ldots, n_{A_{1} \ldots A_{N}})$,
where $n_{k_{1} \ldots k_{N}}$ is the number of fragments
with $k_{j}$ particles of type $j$, $j=1, \ldots, N$.
The set of all such partitions is denoted by $\Pi_{A_{1} \ldots A_{N}}$,
each of which is subject to the constraints
$\sum_{k_{1} \ldots k_{N}} k_{j} n_{k_{1} \ldots k_{N}} = A_{j}$,
$j=1, \ldots, N$.

There are two principle reasons why one-component models are
primarily used in the statistical modeling of nuclear fragmentation.
The first reason is simplicity.
One-component models are fundamentally simpler
than two-component models, and the exact solution for
a class of one-component models has been known for some time.
In addition, as Sneppen~\cite{Sneppen} noted, the
handling of the full two-component system increases the number of
allowable final states dramatically (see
table~\ref{tab:partition-table}).  This makes it impractical to model
these systems by standard Monte Carlo techniques.
As these methods have been the primary technique used in
evaluating realistic models which do not admit exact solutions,
this discourages approaching the problem this way.

The second reason is sufficiency.
A one-component model can always model the physics if the underlying
isotopic states are inaccessible to the observer.
For example, typically only the charge distribution of the fragments
is measured in a fragmentation experiment.
The isotopic distribution for any particular charge $Z$ is not recorded.
So a one-component model distinguishing protons is
sufficient to model the observables measured.
The weight for a particular charge fragment distribution is then given by
summing the two-component weights of all the states which have an
equivalent configuration of protons.  Since this weight
is a complicated sum of two-component weights, there
is no guarantee that the one-component weight will be simple.
Indeed, this is usually not the case, as we will discover in
section~\ref{sec:Isotopic-equivalence}.

However, there are good reasons to prefer modeling nuclear systems
with a two-component model, especially if such a model
is amenable to analytic solution.
For instance, any choice of statistical weight
based on the fundamental statistics of
the constituent particles must by necessity be based on a
two-component model.
Also, unlike one-component models, definite predictions about
the isotopic distribution of fragments could be made for a
two-component model.
Additionally, nuclear effects such as proton-neutron symmetry,
pairing and Coulomb interactions could be modeled, effects which are
difficult (if not impossible) to treat in a one-component model.

This paper shows that a particular class of
two-component models are in fact exactly solvable by recursion.
Realistic models lie within this class, and their solution
obviates the need for Monte Carlo and allows for the exact evaluation of
all expectation values.  A discussion of the relation of partitions
to the permutation group $S_{A}$ provides a closed form solution for
a specific two-component model, as well as a proof that all the
one-component models of the same form as those in ref.~\cite{Chase}
are equivalent to a particular two-component model.

\section{Isotopic models and their solution}
\label{sec:Isotopic-models}

Earlier papers~\cite{Mekjian1,Mekjian2,Mekjian3,Chase}
discussed the one-component model where each state
$\vec{n} \in \Pi_{Z}$ was given the weight
\begin{equation}
P_{Z}(\vec{n}, \vec{x})
  = {W_{Z}(\vec{n}, \vec{x}) \over Z_{Z}(\vec{x})}
  = {1 \over Z_{Z}(\vec{x})} \prod_{j=1}^{Z} {x_{j}^{n_{j}} \over n_{j}!}
\label{eq:one-component-weight}
\end{equation}
The partition functions can be obtained by recursion,
\begin{equation}
Z_{Z}(\vec{x})
  = {1 \over Z} \sum_{j=1}^{Z} j x_{j} Z_{Z-j}(\vec{x})
\label{eq:one-component-partition-function}
\end{equation}
and all the ensemble averages are simple functions of the
partition function.

In analogy with the one-component model, we can
define a canonical weight for $\vec{n} \in \Pi_{ZN}$ by
\begin{equation}
P_{ZN}(\vec{n}, \vec{x})
  = {W_{ZN}(\vec{n}, \vec{x}) \over Z_{ZN}(\vec{x})}
  = {1 \over Z_{ZN}(\vec{x})}
    \prod_{j k} {x_{j k}^{n_{j k}} \over n_{j k}!}
\label{eq:two-component-weight}
\end{equation}
where $\vec{x}$ is a parameter vector which determines the
fragmentation behavior.  The $1/n_{j k}!$ are the
expected Gibbs factors which arise from the
exchange statistics of identical fragments.
The solution of this system can be derived from the following
property of the partition function:
\begin{equation}
{\partial Z_{ZN} \over \partial x_{j k}} = Z_{Z-j, N-k}(\vec{x})
\end{equation}
where $Z_{j k}(\vec{x}) = 0$ when $j<0$ or $k<0$.

{}From this property any moment can be calculated in terms of the
partition functions.  For example,
\begin{equation}
\langle n_{j k} \rangle
  = {x_{j k} \over Z_{Z N}(\vec{x})}
      {\partial Z_{Z N} \over \partial x_{j k}}
  = x_{j k} {Z_{Z-j, N-k}(\vec{x}) \over Z_{ZN}(\vec{x})}
\label{eq:expectation-njk}
\end{equation}
or in general,
\begin{eqnarray}
\left\langle \prod_{j k} [n_{j k}]_{p_{j k}} \right\rangle
  & = & {1 \over Z_{Z N}(\vec{x})}
        \prod_{j k} x_{j k}^{p_{j k}}
        {\partial^{p} Z_{Z N} \over
        \partial x_{1 0}^{p_{1 0}} \ldots \partial x_{Z N}^{p_{Z N}}}
        \nonumber \\
  & = & \prod_{j k} x_{j k}^{p_{j k}}
        {Z_{Z-\sum j p_{j k}, N-\sum k p_{j k}}(\vec{x}) \over
        Z_{Z N}(\vec{x})}
\end{eqnarray}
where $\vec{p}$ is a vector of integers, $p = \sum_{j k} p_{j k}$
and the falling factorial $[z]_{k}$ is given by
$[z]_{k} = (z-k+1)[z]_{k-1}$, with $[z]_{0} = 1$.

Of course, the partition functions are needed to compute these
expressions.  The constraints $\sum_{j k} j n_{j k} = Z$,
$\sum_{j k} k n_{j k} = N$ hold for all configurations,
so they must also hold for the expectation
values.  This allows us to express the partition function as a
recursively defined function:
\begin{eqnarray}
Z_{ZN}(\vec{x})
  & = & {1 \over Z} \sum_{j k} j x_{j k} Z_{Z-j, N-k}(\vec{x}) \nonumber \\
Z_{ZN}(\vec{x})
  & = & {1 \over N} \sum_{j k} k x_{j k} Z_{Z-j, N-k}(\vec{x}) \;.
\label{eq:two-component-partition-function}
\end{eqnarray}
Since $Z_{0 0}(\vec{x}) = 1$, these two relations can be used to
iteratively construct the set of partition functions for any
parameter vector $\vec{x}$.

It should be noted that the technique
used to solve this model applies equally well to any number of
components.  The $N$-component system of $A=\sum_{j=1}^{N} A_{j}$
particles with the weight
\begin{equation}
P_{A_{1} \ldots A_{N}}(\vec{n}, \vec{x})
  = {1 \over Z_{A_{1} \ldots A_{N}}(\vec{x})}
    \prod_{k_{1} \ldots k_{N}}
    {x_{k_{1} \ldots k_{N}}^{n_{k_{1} \ldots k_{N}}} \over
      n_{k_{1} \ldots k_{N}}!}
\label{eq:N-component-weight}
\end{equation}
has a recursively defined set of partition functions
determined by the $N$ equations
\begin{equation}
Z_{A_{1} \ldots A_{N}}(\vec{x}) = {1 \over A_{j}}
  \sum_{k_{1} \ldots k_{N}} k_{j} x_{k_{1} \ldots k_{N}}
  Z_{A_{1}-k_{1}, \ldots, A_{N}-k_{N}}(\vec{x})
\label{N-component-partition-function}
\end{equation}
for $j=1, \ldots, N$ where $Z_{00 \ldots 00}(\vec{x}) = 1$.
All the ensemble averages can be determined solely
from these partition functions and
the parameter vector $\vec{x}$.

\section{Relation to the Permutation group}
\label{sec:Permutation-group}

It is useful to consider an alternative representation of these
models.  Consider the set of all permutations on $A$ objects,
the symmetric group $S_{A}$.
Typically a permutation $p \in S_{A}$ is represented by
a vector of integers $\vec{p} = (p_{1}, \ldots, p_{A})$
where $p_{j}$ is the result of the permutation acting on $j$,
i.e. $j \stackrel{p}{\rightarrow} p_{j}$.  We can also represent $p$
by its cycle decomposition.  In this representation,
the permutation is a set of disjoint cycles of varying length.
Each cycle contains all the numbers which permute among
themselves under the action of the permutation.

If we identify each element in the permutation vector with
a particle, we can consider a cycle in the cycle decomposition
of a permutation as representing a cluster of particles.
This defines a map from the set of
permutations $S_{A}$ onto the set of partitions $\Pi_{A}$ by cycle
decomposition.
For example, table~\ref{tab:permutation-table} details this for $A=4$.
The number of permutations which map to a particular
partition of $A$ was computed by Cauchy, and is given by
\begin{equation}
M_{2}(\vec{n}) = A! \prod_{k=1}^{A} {1 \over n_{k}! k^{n_{k}}}
\end{equation}
Since this is of the same form as the weight given by
the one-component partition model, we see that the model
given by the permutation weight
\begin{equation}
P_{A}(p, \vec{z})
  = {1 \over Z_{A}(\vec{z})}
    \prod_{j=1}^{A} z_{j}^{n_{j}(p)}
\label{eq:one-component-permutation-weight}
\end{equation}
is equivalent to the partition model given by
eq.~(\ref{eq:one-component-weight}) with $x_{j}$ given by
\begin{equation}
x_{j} = {z_{j} \over j}
\end{equation}

We can also map the set of permutations $S_{A}$ onto the
set of partitions $\Pi_{ZN}$ where $A=Z+N$ in an analogous way.
Identify elements $1, \ldots, Z$ in the permutation as protons,
and elements $Z+1, \ldots, Z+N$ as neutrons.  Each cycle in the
permutation is mapped to a cluster by counting the
number of ``neutrons'' and ``protons'' in that particular cycle.
As an example, this is done for $Z=2$, $N=2$ in
table~\ref{tab:permutation-table}.
The number of permutations which map in this way
to a particular $\vec{n} \in \Pi_{ZN}$
is given by the two-component generalization of Cauchy's formula
\begin{equation}
M_{2}(\vec{n}) = Z! N! \prod_{j k} {1 \over n_{j k}!}
  \left( {(j+k-1)! \over j! k!} \right)^{n_{j k}}
\end{equation}
which satisfies the sum rule
$\sum_{\vec{n} \in \Pi_{ZN}} M_{2}(\vec{n}) = (Z+N)!$.
Since this number is of the same form as
eq.~(\ref{eq:two-component-weight}),
the permutation model given by
\begin{equation}
P_{ZN}(p, \vec{z})
  = {1 \over Z_{ZN}(\vec{z})} \prod_{j k} z_{j k}^{n_{j k}(p)}
\label{eq:two-component-permutation-weight}
\end{equation}
is equivalent to the model given by eq.~(\ref{eq:two-component-weight}) with
\begin{equation}
x_{j k} = z_{j k} {(j+k-1)! \over j! k!}
\label{eq:two-component-connection}
\end{equation}

This can be generalized to any number of components.
The $N$-component generalization of Cauchy's
number for $A=\sum_{j=1}^{N} A_{j}$ is given by
\begin{equation}
M_{2}(\vec{n}) = \prod_{j=1}^{N} A_{j}!
  \prod_{k_{1} \ldots k_{N}} {1 \over n_{k_{1} \ldots k_{N}}!}
  \left(
    {(\sum_{j} k_{j} - 1)! \over k_{1}! \cdots k_{N}!}
  \right)^{n_{k_{1} \ldots k_{N}}}
\end{equation}
and the permutation model
\begin{equation}
P_{A_{1} \ldots A_{N}}(\vec{z})
  = {1 \over Z_{A_{1} \ldots A_{N}}(\vec{z})}
    \prod_{k_{1} \ldots k_{N}}
    z_{k_{1} \ldots k_{N}}^{n_{k_{1} \ldots k_{N}}(p)}
\label{eq:N-component-permutation-weight}
\end{equation}
is equivalent to the partition model given by
eq.~(\ref{eq:N-component-weight}) with $x_{k_{1} \ldots k_{N}}$ given by
\begin{equation}
x_{k_{1} \ldots k_{N}}
  = z_{k_{1} \ldots k_{N}}
    {(\sum_{j} k_{j} - 1)! \over k_{1}! \cdots k_{N}!}
\end{equation}

Using the two-component model as defined over the permutation
group allows us to construct a closed-form solution for the
partition function of a particular model.
Consider eq.~(\ref{eq:two-component-permutation-weight}) with
$z_{j k} = x$, where $x$ is a parameter characterizing the degree of
fragmentation of the system.
Each permutation contributes $x^{m}$ to the partition function,
and there are $|S_{A}^{(m)}|$ permutations with $m$ fragments,
where $S_{A}^{(m)}$ is the Stirling number of the first kind.
Therefore $Z_{ZN}(\vec{z}) = x(x+1) \ldots (x+A-1)$.
This implies that the partition model with
\begin{equation}
x_{j k} = x {(j+k-1)! \over j! k!} = {x \over j+k} {j+k \choose j}
\label{eq:two-component-ewens-model}
\end{equation}
has its partition function given by the closed-form expression
\begin{equation}
Z_{Z N}(x)
  = {1 \over Z! N!} {\Gamma(x+Z+N) \over \Gamma(x)}
\label{eq:two-component-ewens-partition-function}
\end{equation}
The expected number of fragments with $j$ protons
and $k$ neutrons is
\begin{equation}
\langle n_{j k} \rangle
  = x {Z \choose j} {N \choose k} B(j+k,x+A-j-k)
\label{eq:two-component-ewens-njk}
\end{equation}
where $B(x,y)$ is the Euler Beta function,
$B(x,y) = \Gamma(x) \Gamma(y)/ \Gamma(x+y)$.
The expected number of fragments of charge $j$,
$n_{j} = \sum_{k=0}^{N} n_{j k}$ is given by
\begin{equation}
\langle n_{j} \rangle
  = x {Z \choose j} B(j,x+Z-j)
\label{eq:two-component-ewens-nj}
\end{equation}
which can be shown by applying the N\"orlund formula
$[x+y]^{n} = \sum_{k=0}^{n} {n \choose k} [x]^{k} [y]^{n-k}$
to eq.~(\ref{eq:two-component-ewens-njk}),
where the rising factorial is given by $[x]^{k} = (x+k-1) [x]^{k-1}$ and
$[x]^{0} = 1$.

Notice that eq.~(\ref{eq:two-component-ewens-nj})
is identical to the one-component expectation value for
the model given by eq.~(\ref{eq:one-component-weight})
with $x_{j} = x/j$ (see \cite{Mekjian3}).  This is not a coincidence.
In fact, the relation of the partition models to the permutation
group allows us to prove in the next section that there exist
two-component models of the form of eq.~(\ref{eq:two-component-weight})
for every one-component model of the form of
eq.~(\ref{eq:one-component-weight}) such that the results of the
one-component model are preserved when a summation over isotopes
is performed.  In this case, the two-component model given by
eq.~(\ref{eq:two-component-ewens-model}) is equivalent to the
one-component model given by $x_{j} = x/j$.

\section{Isotopic and Permutation Equivalence}
\label{sec:Isotopic-equivalence}

As mentioned in section~\ref{sec:Permutation-group},
two-component models are sometimes equivalent to
one-component models.  In this section, we make this notion precise
as well as show under what conditions this is true.

A two-component model is said to be
{\em isotopically equivalent} to a one-component model if for every
function $f(\vec{n})$, the one-component ensemble average
$\langle f \rangle_{Z} =
\sum_{\vec{n}} f(\vec{n}) P_{Z}(\vec{n})$ is equal to
the two-component ensemble average
$\langle f  \rangle_{ZN} =
\sum_{\vec{n}'} f(\vec{n}(\vec{n}')) P'_{ZN}(\vec{n}')$
where $n_{j}(\vec{n}') = \sum_{k=0}^{N} n'_{j k}$.
For example,
$\langle n_{j} \rangle_{Z} = \sum_{k=0}^{N} \langle n_{j k}
\rangle_{ZN}$
must hold if two models are considered to be isotopically
equivalent.  As mentioned in the introduction, this is true if and only if
\begin{equation}
P_{Z}(\vec{n}) = \sum_{\vec{n}' \in \Pi_{ZN}(\vec{n})} P'_{ZN}(\vec{n}')
\label{eq:isotopic-equivalence}
\end{equation}
where $\Pi_{ZN}(\vec{n})$ denotes all the partition vectors
$\vec{n}' \in \Pi_{ZN}$ such that $n_{j} = \sum_{k=0}^{N} n'_{j k}$.

If we consider deriving the models over the permutation group, we
can define the notion of {\em permutation equivalence}
of $P'_{Z'N'}(p')$ to $P_{ZN}(p)$ by the condition
\begin{equation}
P_{ZN}(p) = \sum_{p' \in S_{Z'+N'}(p)} P'_{Z'N'}(p')
\label{eq:permutation-equivalence}
\end{equation}
where $Z' \ge Z$, $N' \ge N$ and $S_{Z'+N'}(p)$ is the set of all
permutations $p' \in S_{Z'+N'}$ which give $p$ by the following
construction.  Decompose $p'$ into its cycle representation.
Eliminate the numbers $Z+1, \ldots, Z'$ and $Z'+N+1, \ldots, Z'+N'$
from the cycles.  Renumber the elements $Z'+1, \ldots, Z'+N$
as $Z+1, \ldots, Z+N$ to recover the permutation in $S_{Z+N}$.
In other words, if $k$ is an element to eliminate, find $j$ such that
$p_{j} = k$, and modify the permutation vector so that $p_{j} = p_{k}$.
This removes the element from the permutation.
Do this for all the numbers to eliminate, resulting in a
new permutation isomorphic to a permutation in $S_{Z+N}$.
For example, $p' = (1,5)(3)(4,6,2) \in S_{6}$ becomes
$(1,5)(4,2)$ by eliminating $3,6$.  After renumbering this becomes
$p = (1,4)(2,3) \in S_{4}$.

Permutation equivalence of $P'_{ZN}(p')$ to $P_{Z0}(p)$ implies
isotopic equivalence of $M_{2}(\vec{n}'(p')) P'_{ZN}(\vec{n}'(p'))$ to
$M_{2}(\vec{n}(p)) P_{Z0}(\vec{n}(p))$, assuming the permutation
weights are only functions of $\vec{n}(p)$, i.e. all
permutations which have the same partition configuration
have the same weight.  One particular permutation
weight which satisfies this condition is the unnormalized weight
$W_{ZN}(p, \vec{z}) = \prod_{j k} z_{j k}^{n_{j k}(p)}$.
We will now show that this weight with $A=Z+N$ is permutationally
equivalent to the $Z, N+1$ weight with the same $z_{j k}$ if
$z_{j k} = z_{j}$.  This implies by induction that the $Z0$ model is
permutationally equivalent to the $ZN$ model, and therefore that the two
weights corrected by the Cauchy factor are isotopically equivalent.

For any permutation $p \in S_{A}$,
there are $A+1$ permutations in $S_{A+1}(p)$.  One of
these permutations has the element $A+1$ in its own cycle.
It has weight $W_{Z, N+1}(p', \vec{z}) = W_{ZN}(p, \vec{z}) z_{01}$.
$A$ of the permutations are hybrids, with the extra element combined
with a cycle from $p$.  Suppose the element is in a cycle
of $j$ neutrons, $k$ protons.  The weight of that cycle is
$W_{Z, N+1}(p', \vec{z}) = W_{ZN}(p, \vec{z}) z_{j, k+1}/z_{j k}$,
since the introduction of the additional element has reduced
$n_{j k}$ by one, but increased $n_{j, k+1}$ by one.  There are $j+k$
possible places for the element to be inserted into that cycle, and
there are $n_{j k}$ cycles of that type it can combine with.  So
the following is true
\begin{equation}
\sum_{p' \in S_{A+1}(p)} {W_{Z, N+1}(p') \over W_{Z N}(p)}
  = z_{01} + \sum_{j k} f_{j k} n_{j k}
\label{eq:step-in-proof}
\end{equation}
where $f_{j k} = (j+k) z_{j, k+1}/z_{j k}$.
If $z_{j k}$ is independent of $k$, this reduces to $z_{0 1}+N+Z$.
In this case, we can multiply both sides of the equation by
$W_{Z N}(p)$ and sum over $\forall p \in S_{A}$ to arrive at
\begin{equation}
Z_{Z, N+1}(\vec{z}) = Z_{Z N}(\vec{z}) (z_{0 1}+Z+N) \;.
\end{equation}
This implies eq.~(\ref{eq:permutation-equivalence}) when substituted
back into eq.~(\ref{eq:step-in-proof}).  So under these conditions,
$P_{Z N}$ is equivalent to $P_{Z, N+1}$.  Therefore, $P_{Z 0}$ is
equivalent to $P_{Z N}$ by induction and transitivity,
and eq.~(\ref{eq:one-component-weight}) is equivalent to
eq.~(\ref{eq:two-component-weight}) for $x_{j k}$ given by
\begin{equation}
x_{j k} = x_{j} {j+k-1 \choose k}
\label{eq:equivalent-xjk}
\end{equation}
The partition function for this two-component model is
\begin{equation}
Z_{Z N}(\vec{x})
  = {\Gamma(z_{0 1}+Z+N) \over \Gamma(z_{0 1}+Z)} Z_{Z}(\vec{x})
\label{eq:equivalent-partition-function}
\end{equation}

Notice that at no point was it asserted that these are the
only two-component models that are isotopically equivalent to
the one-component models.
In the previous discussion, the key property was
the independence of $\sum_{j k} f_{j k} n_{j k}$ from $\vec{n}$.
Perhaps we could satisfy this
with choices other than $f_{j k} = j + k$.
In fact, it can be proven that this condition is satisfied only by
$f_{j k} = \alpha {A j / Z} + (1 - \alpha) {A k / N}$
where $\alpha$ is arbitrary.  However, $\alpha \in [0,1]$ to
guarantee nonnegative probabilities in the statistical model.
Given $z_{j 0}$, we could then construct permutationally equivalent
two-component models specified by $\alpha$ and the recursion relation
\begin{equation}
z_{j, k+1}
  = z_{j k} {A \over j+k} \left(
    \alpha {j \over Z} + (1 - \alpha) {k \over N}
  \right)
\label{eq:equivalent-zjk}
\end{equation}
and evaluate the equivalent partition weights using
eqs.~(\ref{eq:two-component-weight}),~(\ref{eq:two-component-connection}).

If we require that the models be isotopically equivalent for
any choice of $N$, these are the only models given by
eqs.~(\ref{eq:one-component-weight}),~(\ref{eq:two-component-weight})
that are isotopically equivalent.
This is true because isotopic equivalence for any
$N$ implies permutation equivalence of $N$ to $N+1$.
Permutation equivalence requires $\sum_{j k} f_{j k} n_{j k}$ to be
independent of $\vec{n}$, which can be shown to be true only if
$z_{j k}$ is of the form given above.
Notice that this condition is very similar to Kingman's
non-interference condition on
one-component partition weights \cite{Kingman}.
We require that the neutrons do not ``interfere'' with the
overall proton distribution, just as in one-component models Kingman
required that the addition of new objects does not interfere with the
probability structure of the original set of objects.

This technique can be generalized to construct an $N+1$-component
model that is equivalent to an $N$-component model.  For example, if the
$N$-component weight is given by eq.~(\ref{eq:N-component-weight}),
then an equivalent $N+1$-component model is given by the parameter vector
defined by
\begin{equation}
x_{k_{1} \ldots k_{N+1}}
  = x_{k_{1} \ldots k_{N}} {\sum_{j=1}^{N+1} k_{j} - 1 \choose k_{N+1}}
\end{equation}

These results provide a simple way of constructing two-component models
from the one-component models explored in earlier papers.
For example, one could apply
eqs.~(\ref{eq:equivalent-xjk}),~(\ref{eq:equivalent-partition-function})
to the chain model, $x_{j} = x$, to construct a two-component
extension of the chain model as developed by Gross, et.
al.~\cite{Gross1}.

\section{Application to Nuclear Fragmentation}
\label{sec:Application-to-nuclear}

The two-component analogs of one-component models
given by eq.~(\ref{eq:equivalent-zjk}) could be taken
as the starting point for a fully isotopic model of
fragmentation.  These models would contain no
surprises in their one-component expectation values, and would
provide predictions for the isotopic behavior.

This is perhaps too strong a requirement.
Equation~(\ref{eq:equivalent-zjk}) constrains the
set of possible partition weights, forbidding a number
of physically interesting choices for $x_{j k}$.  For example,
if the symmetric term $\exp \{ -\alpha_{s} (j-k)^{2}/(j+k) \}$ appeared in
the parameter $z_{j k}$, the resulting parameter vector could never
satisfy eq.~(\ref{eq:equivalent-zjk}).  Since such a symmetry term should
appear due to the symmetry term in the binding energy of nuclei, we
should not expect
realistic models to be isotopically equivalent to the models
considered in earlier papers.

With that in mind, we start with a simplified model of
nuclear fragmentation specified by the following choice for $x_{j k}$
\begin{equation}
x_{j k} = {x \over (j+k)^{\tau}} {j+k \choose j}
  \exp \left\{ -{\alpha_{s} (j-k)^2 \over j+k} \right\} \;.
\end{equation}
We will find it appropriate sometimes to modify this when $j+k = 1$,
i.e. to avoid the exponential suppression factor $e^{-\alpha_{s}}$
for monomers.
Let us specialize to the case $\tau=1$ and consider the limiting cases.
When $\alpha_{s} \rightarrow 0$, the model becomes the two-component
analog of the model discussed in
section~\ref{sec:Permutation-group}.  The partition function
has a closed-form solution, which is given by
eq.~(\ref{eq:two-component-ewens-partition-function}).
When $\alpha_{s} \rightarrow \infty$, $x_{j k}$ is proportional to a Kronecker
delta function and the model reduces to the
one-component model specified by parameters
\begin{equation}
x_{k} = {x \over 2k} {2k \choose k}
\end{equation}
where $k$ indexes both proton and neutron number,
since only symmetric fragments with $j=k$ are allowed.
In this case, the partition function also has a closed-form solution
\begin{equation}
Z_{A}(x) = {x \over A!} {\Gamma(x+2A) \over \Gamma(x+A+1)}
\end{equation}
As an aside, note that all models with
$x_{k} = {x \over nk} {nk \choose k}$ for $n=1,2,3,\ldots$
have closed-form solutions
given by $Z_{A}(x) = (x/A!) (\Gamma(x+nA)/\Gamma(x+(n-1)A+1))$.
Models with intermediate values for $\alpha_{s}$ are not as simple as
these two limits, but can still be computed easily.
Figures~\ref{fig:sym-distribution},~\ref{fig:asym-distribution}
show the prediction of this model at $x=1$ and various $\alpha_{s}$
for a symmetric case $Z=N=50$ as well as an asymmetric case $Z=50$,
$N=75$.  These figures display the total charge distribution
$\langle n_{j} \rangle = \sum_{k=0}^{N} \langle n_{j k} \rangle$,
as well as the distribution of isotopes for a
particular charge, specifically carbon, i.e.
$\langle n_{j k} \rangle$ for $j=6$.

{}From the total charge distribution figures,
we see that increasing $\alpha_{s}$ increases the
number of large fragments.  This is due to the exponential suppression
of asymmetric fragments, which make up a large fraction of the
partition space, especially for smaller fragments.
The isotopic distribution figures reveal that the distribution of isotopes is
essentially Gaussian.
As seen earlier~\cite{Sneppen}, an initial
asymmetry in the relative numbers of protons and neutrons yields a
similar asymmetry in the distribution of fragments.
This asymmetry can be diminished by increasing $\alpha_{s}$ and removing
the exponential suppression from monomer parameters ($x_{01}$, $x_{10}$),
as was done in the figures.
Indeed, when $\alpha_{s} \rightarrow \infty$, such a model would
allow only symmetric fragments and monomers.

To this simple model, we can add other effects, eventually
including all the terms in the semi-empirical mass-formula.
For example, nucleon pairing interactions can be added
using the term $\exp \{ \alpha_{p} p_{j k} / (j+k) \}$ where
$p_{j k}$ is $0$ for even-odd nuclei, $-1$ for odd-odd nuclei, and
$+1$ for even-even nuclei.  This can be done for all intrafragment
interactions.  Interfragment interactions do not fit the form of
eq.~(\ref{eq:two-component-weight}) and must be handled in some kind
of mean-field manner in this approach.  Coulomb interactions are the
only persistent long-range interactions in the thermalized phase.
Suppose each fragment interacts with all the others via the Coulomb
interaction.  This contributes an interaction energy given by
\begin{equation}
E_{I} = {1 \over 2} \sum_{j k l m} n_{j k}
	(n_{l m} - \delta_{j l} \delta_{k m}) E_{j k l m}^{I}
\end{equation}
If we choose $E_{j k l m}^{I} = j l e^2 / \langle r \rangle$ for the
Coulomb interaction, then
\begin{equation}
E_{I} = {e^{2} \over 2 \langle r \rangle} (Z^2 - \sum_{j k} j^2 n_{j k})
\end{equation}
which suggests we include interfragment Coulomb interactions by appending
the term $\exp (\alpha_{c} j^2)$ to $x_{j k}$.  We can estimate
$\alpha_{c}$ by assuming
$\langle r \rangle \approx V^{1/3} \approx r_{0} A^{1/3}$.

We postpone investigating such a realistic choice for $x_{j k}$ for
a later paper.

\section{Conclusion and Summary}
\label{sec:Conclusion}

This paper established a non-trivial exactly solvable isotopic
model of fragmentation.
The model is specificed by a set of parameters $x_{j k}$, where
$x_{j k}$ includes all the physical parameters pertinent to the
fragmenting behavior of clusters of $j$ protons, $k$ neutrons.
This model is related to the permutation group and as a result,
it was proven that for certain choices of $x_{j k}$ it directly
reduces to a one-component model when a summation over isotopes is
performed.
The usefulness of this model in nuclear fragmentation is manifest, and
a simplified model analyzed here contains most of the salient features of
the phenomenon.  A more realistic model is also discussed.

\acknowledgments
This work supported in part by the National Science Foundation
Grant \# NSFPHY 92-12016. One author (K.C.) wishes to thank Rutgers
University Excellence Graduate Fellowship for providing support during
part of this research.

\begin{figure}
\caption{Overall (a) and Isotopic (b) distribution of
fragments for $Z=50$, $N=50$,
$x=1$, $\tau=1$ and various $\alpha_{s}$.
Isotopic distribution is for carbon, $Z=6$.}
\label{fig:sym-distribution}
\end{figure}

\begin{figure}
\caption{Overall (a) and Isotopic (b) distribution of
fragments for $Z=50$, $N=75$,
$x=1$, $\tau=1$ and various $\alpha_{s}$.
Isotopic distribution is for carbon, $Z=6$.}
\label{fig:asym-distribution}
\end{figure}

\begin{table}
\caption{Number of partitions for $Z$ protons and $N$ neutrons.}
\label{tab:partition-table}
\begin{tabular}{rrrrr}
$Z$ & $N$ & $P_{Z}$ & $P_{N+Z}$ &                $P_{ZN}$ \\
10 & 10 &      42 &         627 &                   59521 \\
20 & 20 &     627 &       37338 &              3026 73029 \\
30 & 35 &    5604 &    20 12558 &         231 50477 43167 \\
40 & 50 &   37338 &   566 34173 &     6 00944 62493 61633 \\
50 & 70 & 9 66467 & 18443 49560 & 26953 04447 12166 27712
\end{tabular}
\end{table}

\begin{table}
\caption{The twenty-four permutations of $S_{4}$ and their
partition representations.  The first two columns give the typical
representation and cycle decomposition of the permutations.
The third column gives the equivalent one-component partition vectors
$\vec{n} = (n_{1}, n_{2}, n_{3}, n_{4})$.  The last column gives the
equivalent two-component partition vectors
$\vec{n} = (n_{01}, n_{02}, n_{10}, n_{11}, n_{12}, n_{20}, n_{21}, n_{22})$
for $Z=2$, $N=2$.}
\label{tab:permutation-table}
\begin{tabular}{llll}
$\vec{p}$ & $p$ & $\vec{n}_{4}$ & $\vec{n}_{22}$ \\
(1,2,3,4) & (1)(2)(3)(4) & (4,0,0,0) & (2,0,2,0,0,0,0,0) \\
(1,2,4,3) & (1)(2)(3,4)  & (2,1,0,0) & (0,0,2,0,0,1,0,0) \\
(1,3,2,4) & (1)(2,3)(4)  & (2,1,0,0) & (1,0,1,1,0,0,0,0) \\
(1,3,4,2) & (1)(2,3,4)   & (1,0,1,0) & (0,0,1,0,1,0,0,0) \\
(1,4,2,3) & (1)(2,4,3)   & (1,0,1,0) & (0,0,1,0,1,0,0,0) \\
(1,4,3,2) & (1)(2,4)(3)  & (2,1,0,0) & (1,0,1,1,0,0,0,0) \\
(2,1,3,4) & (1,2)(3)(4)  & (2,1,0,0) & (2,1,0,0,0,0,0,0) \\
(2,1,4,3) & (1,2)(3,4)   & (0,2,0,0) & (0,1,0,0,0,1,0,0) \\
(2,3,1,4) & (1,2,3)(4)   & (1,0,1,0) & (1,0,0,0,0,0,1,0) \\
(2,3,4,1) & (1,2,3,4)    & (0,0,0,1) & (0,0,0,0,0,0,0,1) \\
(2,4,1,3) & (1,2,4,3)    & (0,0,0,1) & (0,0,0,0,0,0,0,1) \\
(2,4,3,1) & (1,2,4)(3)   & (1,0,1,0) & (1,0,0,0,0,0,1,0) \\
(3,1,2,4) & (1,3,2)(4)   & (1,0,1,0) & (1,0,0,0,0,0,1,0) \\
(3,1,4,2) & (1,3,4,2)    & (0,0,0,1) & (0,0,0,0,0,0,0,1) \\
(3,2,1,4) & (1,3)(2)(4)  & (2,1,0,0) & (1,0,1,1,0,0,0,0) \\
(3,2,4,1) & (1,3,4)(2)   & (0,0,0,1) & (0,0,0,0,0,0,0,1) \\
(3,4,1,2) & (1,3)(2,4)   & (0,2,0,0) & (0,0,0,2,0,0,0,0) \\
(3,4,2,1) & (1,3,2,4)    & (0,0,0,1) & (0,0,0,0,0,0,0,1) \\
(4,1,2,3) & (1,4,3,2)    & (0,0,0,1) & (0,0,0,0,0,0,0,1) \\
(4,1,3,2) & (1,4,2)(3)   & (1,0,1,0) & (1,0,0,0,0,0,1,0) \\
(4,2,1,3) & (1,4,3)(2)   & (1,0,1,0) & (0,0,1,0,1,0,0,0) \\
(4,2,3,1) & (1,4)(2)(3)  & (2,1,0,0) & (1,0,1,1,0,0,0,0) \\
(4,3,1,2) & (1,4,2,3)    & (0,0,0,1) & (0,0,0,0,0,0,0,1) \\
(4,3,2,1) & (1,4)(2,3)   & (0,2,0,0) & (0,0,0,2,0,0,0,0)
\end{tabular}
\end{table}

\end{document}